\documentclass[lettersize,journal]{IEEEtran}
\usepackage{amsmath,amsfonts}
\usepackage{algorithmic}
\usepackage{algorithm}
\usepackage{array}
\usepackage[caption=false,font=normalsize,labelfont=sf,textfont=sf]{subfig}
\usepackage{textcomp}
\usepackage{stfloats}
\usepackage{url}
\usepackage{verbatim}
\usepackage{graphicx}
\usepackage{cite}
\usepackage{color}
\begin{document}

\title{Interleaved Transceiver Design for a Continuous- Transmission MIMO-OFDM ISAC System}

{\author{Yating Chen,~\IEEEmembership{IEEE Student Member}, Cai Wen,~\IEEEmembership{IEEE Member},  Yan Huang,~\IEEEmembership{IEEE Member}, \\
Jinye Peng, Wei Hong,~\IEEEmembership{IEEE Fellow}, Timothy N. Davidson,~\IEEEmembership{IEEE Fellow}

\thanks{(Corresponding author: Cai Wen)
Yating Chen, Yan Huang, and Wei Hong are with the State Key Laboratory of Millimeter Waves, School of Information Science and Engineering, Southeast University, Nanjing, China.

Cai Wen and Jinye Peng are with the School of Information Science and Technology, Northwest University, Xi'an, China.

Timothy N. Davidson is with the Department of Electrical and Computer Engineering, McMaster University, Hamilton, Canada.}}

\maketitle

\begin{abstract}
This paper proposes an interleaved transceiver design method for a multiple-input multiple-output (MIMO) integrated sensing and communication (ISAC) system utilizing orthogonal frequency division multiplexing (OFDM) waveforms. We consider a continuous transmission system and focus on the design of the transmission signal and a receiving filter in the time domain for an interleaved transmission architecture. For communication performance, constructive interference (CI) is integrated into the optimization problem. For radar sensing performance, the integrated mainlobe-to-sidelobe ratio (IMSR) of the beampattern is considered to ensure desirable directivity. Additionally, we tackle the challenges of inter-block interference and eliminate the spurious peaks, which are crucial for accurate target detection. Regarding the hardware implementation aspect, the power of each time sample is constrained to manage the peak-to-average power ratio (PAPR). The design problem is addressed using an alternating optimization (AO) framework, with the subproblem for transmitted waveform design being solved via the successive convex approximation (SCA) method. To further enhance computational efficiency, the alternate direction penalty method (ADPM) is employed to solve the subproblems within the SCA iterations. The convergence of ADPM is established, with convergence of the case of more than two auxiliary variables being established for the first time. Numerical simulations validate the effectiveness of our transceiver design in achieving desirable performance in both radar sensing and communication, with the fast algorithm achieving comparable performance with greater computational efficiency.
\end{abstract}

\subsection{Solution of (40)}
\setcounter{equation}{0}
\renewcommand\theequation{A\arabic{equation}}
Suppose that the eigen-decomposition of ${\bf{C}}_{\bf{s}}^H{{\boldsymbol{\Psi }}_\text{SL}}{{\bf{C}}_{\bf{s}}}$ is ${{\bf{Q}}_1}{{\bf{\Lambda }}_1}{\bf{Q}}_1^H$. Since ${{\bf{\Lambda }}_1}$ is diagonal and real, and ${{\bf{Q}}_1}$ is a unitary matrix, the problem in (40) can be transformed into 
\begin{equation}
\label{C1}
    \begin{array}{l}
\mathop {\min }\limits_{{\bf{\hat b}}} \quad \bigl\| {{\bf{\hat b}} - {{\boldsymbol{\varsigma }}^{\left( k \right)}} + {\bf{\hat {\boldsymbol{\mu}} }}_{\bf{b}}^{\left( k \right)}} \bigr\|_2^2\\
\text{s.t. }\quad\;{{{\bf{\hat b}}}^H}{{\bf{\Lambda }}_1}{\bf{\hat b}} - {\mathop{\rm Re}\nolimits} \bigl\{ {{\boldsymbol{\beta }}_1^{\left( j \right)H}{\bf{\hat b}}} \bigr\} \le 0,
\end{array}
\end{equation}
where ${\bf{\hat b}} = {\bf{Q}}_1^H{\bf{b}}$, ${{\bf{\varsigma }}^{\left( k \right)}} = {\bf{Q}}_1^H\left[ {{{\bf{s}}^{\left( k \right)}};{t^{\left( k \right)}}} \right]$, ${\bf{\hat {\boldsymbol{\mu}} }}_{\bf{b}}^{\left( k \right)} = \frac{{{\bf{Q}}_1^H{\boldsymbol{\mu }}_{\bf{b}}^{\left( k \right)}}}{{\rho _1^{\left( k \right)}}}$ and ${\boldsymbol{\beta }}_1^{\left( j \right)H} = \left( {\frac{{2{{\bf{s}}^{\left( j \right)H}}{{\boldsymbol{\Psi }}_\text{ML}}{{\bf{C}}_{\bf{s}}}}}{{{{\bar t}^{\left( j \right)}}}} - \frac{{{{\bf{s}}^{\left( j \right)H}}{{\boldsymbol{\Psi }}_\text{ML}}{{\bf{s}}^{\left( j \right)}}}}{{{{\bar t}^{\left( j \right)2}}}}{\bf{c}}_t^T} \right){{\bf{Q}}_1}$. 
The augmented Lagrangian function is
\begin{equation}
    \begin{aligned}
{\cal{L}}\bigl( {{\bf{\hat b}},\lambda } \bigr) = &\bigl\| {{\bf{\hat b}} - {{\boldsymbol{\varsigma }}^{\left( k \right)}} + {\bf{\hat {\boldsymbol{\mu}} }}_{\bf{b}}^{\left( k \right)}} \bigr\|_2^2  \\
&\quad +\lambda \bigl( {{{{\bf{\hat b}}}^H}{{\bf{\Lambda }}_1}{\bf{\hat b}} - {\mathop{\rm Re}\nolimits} \bigl\{ {{\boldsymbol{\beta }}_1^{\left( j \right)H}{\bf{\hat b}}} \bigr\}} \bigr).
\end{aligned}
\end{equation}

As (\ref{C1}) is a convex problem and we can always find a strictly feasible point for the constraint, the optimal solution can be obtained by solving the KKT conditions: 
\begin{align}
    \lambda \bigl( {{{{\bf{\hat b}}}^H}{{\bf{\Lambda }}_1}{\bf{\hat b}} - {\mathop{\rm Re}\nolimits} \bigl\{ {{\boldsymbol{\beta }}_1^{\left( j \right)H}{\bf{\hat b}}} \bigr\}} \bigr) &= 0,\\
    \lambda  &\ge 0,\\
\label{C5}
    {{\bf{\hat b}}^H}{{\bf{\Lambda }}_1}{\bf{\hat b}} - {\mathop{\rm Re}\nolimits} \bigl\{ {{\boldsymbol{\beta }}_1^{\left( j \right)H}{\bf{\hat b}}} \bigr\} &\le 0,
\end{align}
\begin{equation}
\label{C6}
    {\bf{\hat b}} = {({\bf{I}} + \lambda {{\bf{\Lambda }}_1})^{ - 1}}\bigl( {{{\boldsymbol{\varsigma }}^{\left( k \right)}} - {\bf{\hat {\boldsymbol{\mu}} }}_{\bf{b}}^{\left( k \right)} + \textstyle\frac{1}{2}\lambda {\boldsymbol{\beta }}_1^{\left( j \right)}} \bigr).
\end{equation}
If $\lambda  = 0$, we have ${\bf{\hat b}} = {{\boldsymbol{\varsigma }}^{\left( k \right)}} - {\bf{\hat {\boldsymbol{\mu}} }}_{\bf{b}}^{\left( k \right)}$. If that solution satisfies (\ref{C5}), it is optimal. Otherwise, we have $\lambda  > 0$. The optimal solution in that case can be obtained by solving the following pair of equations
\begin{equation}
\label{C7}
    \left\{ {\begin{array}{*{20}{c}}
{{{{\bf{\hat b}}}^H}{{\bf{\Lambda }}_1}{\bf{\hat b}} - {\mathop{\rm Re}\nolimits} \bigl\{ {{\boldsymbol{\beta }}_1^{\left( j \right)H}{\bf{\hat b}}} \bigr\} = 0},\\
{{\bf{\hat b}} = {{({\bf{I}} + \lambda {{\bf{\Lambda }}_1})}^{ - 1}}\bigl( {{{\boldsymbol{\varsigma }}^{\left( k \right)}} - {\bf{\hat {\boldsymbol{\mu}} }}_{\bf{b}}^{\left( k \right)} + \frac{1}{2}\lambda {\boldsymbol{\beta }}_1^{\left( j \right)}} \bigr)}.
\end{array}} \right.
\end{equation}
According to (\ref{C7}), we have that
\begin{equation}
   \begin{array}{l}
f\left( \lambda  \right) = {\sum\limits_{n = 1}^{{N_\text{s}}{N_\text{t}} + 1} {{a_n}\left| {\frac{{{{\boldsymbol{\varsigma }}^{\left( k \right)}}\left[ n \right] - {\bf{\hat {\boldsymbol{\mu}} }}_{\bf{b}}^{\left( k \right)}\left[ n \right] + \frac{1}{2}\lambda {\boldsymbol{\beta }}_1^{\left( j \right)}\left[ n \right]}}{{1 + \lambda {a_n}}}} \right|} ^2}\\
 \qquad - {\mathop{\rm Re}\nolimits} \left\{ {\sum\limits_{n = 1}^{{N_\text{s}}{N_\text{t}} + 1} {{\boldsymbol{\beta }}_1^{\left( j \right)}{{\left[ n \right]}^ * }\frac{{{{\boldsymbol{\varsigma }}^{\left( k \right)}}\left[ n \right] - {\bf{\hat {\boldsymbol{\mu}} }}_{\bf{b}}^{\left( k \right)}\left[ n \right] + \frac{1}{2}\lambda {\boldsymbol{\beta }}_1^{\left( j \right)}\left[ n \right]}}{{1 + \lambda {a_n}}}} } \right\} = 0,
\end{array}
\end{equation}
where ${a_n}$ is the $n$-th eigenvalue. The function $f\left( \lambda  \right)$ is monotonic and the unique solution can be obtained by bi-section or Newton’s method \cite{r1}. With that solution for $\lambda $, we can obtain ${\bf{\hat b}}$ and the corresponding ${\bf{b}}$.  

\subsection{Solution of (42)}
\setcounter{equation}{0}
\renewcommand\theequation{B\arabic{equation}}
This solution can be derived in a similar way to that in the previous subsection. In particular, let the eigen-decomposition of $\eta {\bf{\bar G}}_{{N_\text{s}} + {N_\text{cp}}}^H{{\bf{\bar G}}_{{N_\text{s}} + {N_\text{cp}}}}$ be ${{\bf{Q}}_2}{{\bf{\Lambda }}_2}{\bf{Q}}_2^H$. The problem in (42) can be transformed to 
\begin{equation}
\label{D1}
    \begin{array}{l}
\mathop {\min }\limits_{{\bf{\hat p}}} 
\quad\bigl\| {{\bf{\hat p}} - {{{\bf{\hat s}}}^{\left( k \right)}} + {\bf{\hat {\boldsymbol{\mu}} }}_{\bf{p}}^{\left( k \right)}} \bigr\|_2^2\\
\text{s.t. }\quad \; {{{\bf{\hat p}}}^H}{{\bf{\Lambda }}_2}{\bf{\hat p}} - {\mathop{\rm Re}\nolimits} \bigl\{ {{\boldsymbol{\beta }}_2^{\left( j \right)H}{\bf{\hat p}}} \bigr\} \le 0,
\end{array}
\end{equation}
where ${\bf{\hat p}} = {\bf{Q}}_2^H{\bf{p}}$, ${{\bf{\hat s}}^{\left( k \right)}} = {\bf{Q}}_2^H{{\bf{s}}^{\left( k \right)}}$, ${\bf{\hat {\boldsymbol{\mu}} }}_{\bf{p}}^{\left( k \right)} = \frac{{{\bf{Q}}_2^H{\boldsymbol{\mu }}_{\bf{p}}^{\left( k \right)}}}{{\rho _2^{\left( k \right)}}}$ and ${\boldsymbol{\beta }}_2^{\left( j \right)} = {\bf{Q}}_2^H{\bf{\bar G}}_{{N_\text{s}} + {N_\text{cp}}}^H{{\bf{g}}^{\left( {i + 1} \right) }}$. The augmented Lagrangian function is
\begin{equation}
    \begin{aligned}
{\cal{L}}\left( {{\bf{\hat p}},\lambda } \right) =
\, & \bigl\| {{\bf{\hat p}} - {{{\bf{\hat s}}}^{\left( k \right)}} + {\bf{\hat {\boldsymbol{\mu}} }}_{\bf{p}}^{\left( k \right)}} \bigr\|_2^2  \\
&+ \lambda \bigl( {{{{\bf{\hat p}}}^H}{{\bf{\Lambda }}_2}{\bf{\hat p}} - {\mathop{\rm Re}\nolimits} \bigl\{ {{\boldsymbol{\beta }}_2^{\left( j \right)H}{\bf{\hat p}}} \bigr\}} \bigr).
\end{aligned}
\end{equation}

Since the KKT conditions are necessary and sufficient, we can show that if $\lambda  = 0$, we have ${\bf{p}} = {{\bf{s}}^{\left( k \right)}} - {\boldsymbol{\mu }}_{\bf{p}}^{\left( k \right)}$. If that solution satisfies $    {{\bf{\hat p}}^H}{{\bf{\Lambda }}_2}{\bf{\hat p}} - {\mathop{\rm Re}\nolimits} \bigl\{ {{\boldsymbol{\beta }}_2^{\left( j \right)H}{\bf{\hat p}}} \bigr\} \le 0$, it is optimal. Otherwise, we have $\lambda  > 0$. The optimal solution can be obtained by solving 
\begin{equation}
    \begin{array}{l}
f\left( \lambda  \right) = {\sum\limits_{n = 1}^{{N_\text{s}}{N_\text{t}}} {{a_n}\left| {\frac{{{{{\bf{\hat s}}}^{\left( k \right)}}\left[ n \right] - {\bf{\hat {\boldsymbol{\mu}} }}_{\bf{p}}^{\left( k \right)}\left[ n \right] + \frac{1}{2}\lambda {\boldsymbol{\beta }}_2^{\left( j \right)}\left[ n \right]}}{{1 + \lambda {a_n}}}} \right|} ^2}\\
 - {\mathop{\rm Re}\nolimits} \left\{ {\sum\limits_{n = 1}^{{N_\text{s}}{N_\text{t}}} {{\boldsymbol{\beta }}_2^{\left( j \right)}{{\left[ n \right]}^ * }\frac{{{{{\bf{\hat s}}}^{\left( k \right)}}\left[ n \right] - {\bf{\hat {\boldsymbol{\mu}} }}_{\bf{p}}^{\left( k \right)}\left[ n \right] + \frac{1}{2}\lambda {\boldsymbol{\beta }}_2^{\left( j \right)}\left[ n \right]}}{{1 + \lambda {a_n}}}} } \right\} = 0,
\end{array}
\end{equation}
where ${a_n}$ is the $n$-th eigenvalue. The function $f\left( \lambda  \right)$ is monotonic and the unique solution can be obtained by bi-section or Newton’s method. With that solution for $\lambda $, we can obtain ${\bf{\hat p}}$ and the corresponding ${\bf{p}}$.

\subsection{Derivation of the Closed-form Solution of (44)}
\setcounter{equation}{0}
\renewcommand\theequation{C\arabic{equation}}
The optimization problem can be transformed to
\begin{equation}
\begin{aligned}
\mathop {\min }\limits_{{{\bf{q}}_l}} \text{ }&{\bf{q}}_l^H{{\bf{q}}_l} - {\bf{q}}_l^H{{\bf{s}}^{\left( k \right)}} + \frac{{{\bf{q}}_l^H{\boldsymbol{\mu }}_{{{\bf{q}}_l}}^{\left( k \right)}}}{{\rho _{3,l}^{\left( k \right)}}} - {{\bf{s}}^{\left( k \right)H}}{{\bf{q}}_l} + \frac{{{\boldsymbol{\mu }}_{{{\bf{q}}_l}}^{\left( k \right)H}{{\bf{q}}_l}}}{{\rho _{3,l}^{\left( k \right)}}}\\
\text{s.t. }&- \left[ {{\mathop{\rm Re}\nolimits} \left( {{\bf{\tilde h}}_l^T{{\bf{q}}_l}{e^{ - j\angle {{\bf{s}}_\text{D}}\left[ l \right]}}} \right) - \sqrt {{\Gamma _u}{\sigma ^2}} \left| {{{\bf{s}}_\text{D}}\left[ l \right]} \right|} \right]\tan \phi \\
 &\quad + {\mathop{\rm Im}\nolimits} \left( {{\bf{\tilde h}}_l^T{{\bf{q}}_l}{e^{ - j\angle {{\bf{s}}_\text{D}}\left[ l \right]}}} \right) \le 0,\\
 &- \left[ {{\mathop{\rm Re}\nolimits} \left( {{\bf{\tilde h}}_l^T{{\bf{q}}_l}{e^{ - j\angle {{\bf{s}}_\text{D}}\left[ l \right]}}} \right) - \sqrt {{\Gamma _u}{\sigma ^2}} \left| {{{\bf{s}}_\text{D}}\left[ l \right]} \right|} \right]\tan \phi \\
 &\quad - {\mathop{\rm Im}\nolimits} \left( {{\bf{\tilde h}}_l^T{{\bf{q}}_l}{e^{ - j\angle {{\bf{s}}_\text{D}}\left[ l \right]}}} \right) \le 0.
\end{aligned}
\end{equation}

\noindent The corresponding KKT conditions are given by
\begin{equation}
\label{E3}
    \begin{array}{l}
{\lambda _1}\left[ { - \frac{{\tan \phi  + j}}{2}{e^{ - j\angle {{\bf{s}}_\text{D}}\left[ l \right]}}{\bf{\tilde h}}_l^T{{\bf{q}}_l} - \frac{{\tan \phi  - j}}{2}{e^{j\angle {{\bf{s}}_\text{D}}\left[ l \right]}}{\bf{q}}_l^H{\bf{\tilde h}}_l^ * } \right.\\
\left. { + \sqrt {{\Gamma _u}{\sigma ^2}} \left| {{{\bf{s}}_\text{D}}\left[ l \right]} \right|\tan \phi } \right] = 0,
\end{array}
\end{equation}
\begin{equation}
\label{E4}
    \begin{array}{l}
{\lambda _2}\left[ { - \frac{{\tan \phi  - j}}{2}{e^{ - j\angle {{\bf{s}}_\text{D}}\left[ l \right]}}{\bf{\tilde h}}_l^T{{\bf{q}}_l} - \frac{{\tan \phi  + j}}{2}{e^{j\angle {{\bf{s}}_\text{D}}\left[ l \right]}}{\bf{q}}_l^H{\bf{\tilde h}}_l^ * } \right.\\
\left. { + \sqrt {{\Gamma _u}{\sigma ^2}} \left| {{{\bf{s}}_\text{D}}\left[ l \right]} \right|\tan \phi } \right] = 0,
\end{array}
\end{equation}
\begin{equation}
    {\lambda _1} \ge 0, {\lambda _2} \ge 0,
\end{equation}
\begin{equation}
\label{E7}
    \begin{array}{l}
 - \frac{{\tan \phi  + j}}{2}{e^{ - j\angle {{\bf{s}}_\text{D}}\left[ l \right]}}{\bf{\tilde h}}_l^T{{\bf{q}}_l} - \frac{{\tan \phi  - j}}{2}{e^{j\angle {{\bf{s}}_\text{D}}\left[ l \right]}}{\bf{q}}_l^H{\bf{\tilde h}}_l^ * \\
 + \sqrt {{\Gamma _u}{\sigma ^2}} \left| {{{\bf{s}}_\text{D}}\left[ l \right]} \right|\tan \phi  \le 0,
\end{array}
\end{equation}
\begin{equation}
\label{E8}
    \begin{array}{l}
 - \frac{{\tan \phi  - j}}{2}{e^{ - j\angle {{\bf{s}}_\text{D}}\left[ l \right]}}{\bf{\tilde h}}_l^T{{\bf{q}}_l} - \frac{{\tan \phi  + j}}{2}{e^{j\angle {{\bf{s}}_\text{D}}\left[ l \right]}}{\bf{q}}_l^H{\bf{\tilde h}}_l^ * \\
 + \sqrt {{\Gamma _u}{\sigma ^2}} \left| {{{\bf{s}}_\text{D}}\left[ l \right]} \right|\tan \phi  \le 0,
\end{array}
\end{equation}
\begin{small}
\begin{equation}
\label{E9}
\begin{array}{l}
{{\bf{q}}_l} ={{\bf{s}}^{\left( k \right)}} - \frac{{{\boldsymbol{\mu }}_{{{\bf{q}}_l}}^{\left( k \right)}}}{{\rho _{3,l}^{\left( k \right)}}} + \left( {{\lambda _1}\frac{{\tan \phi  - j}}{2} + {\lambda _2}\frac{{\tan \phi  + j}}{2}} \right){e^{j\angle {{\bf{s}}_\text{D}}\left[ l \right]}}{\bf{\tilde h}}_l^ * .
\end{array}
\end{equation}
\end{small}
Next, four cases are discussed to obtain the closed-form solution. The original problem is strongly convex, so a feasible solution satisfying KKT conditions is the unique optimal solution of the original problem. Hence, the order in which we check each case does not affect the result.

\subsubsection{Case 1}
If ${\lambda _1} = {\lambda _2} = 0$, we have ${{\bf{q}}_l} = {{\bf{s}}^{\left( k \right)}} - \frac{{{\boldsymbol{\mu }}_{{{\bf{q}}_l}}^{\left( k \right)}}}{{\rho _{3,l}^{\left( k \right)}}}$. If this solution satisfies (\ref{E7}) and (\ref{E8}), it is optimal. Otherwise, go to case 2.

\subsubsection{Case 2}
If ${\lambda _1} = 0$ and ${\lambda _2} > 0$, $ {{\bf{q}}_l}$ takes the form
${{\bf{q}}_l} = {{\bf{s}}^{\left( k \right)}} - \frac{{{\boldsymbol{\mu }}_{{{\bf{q}}_l}}^{\left( k \right)}}}{{\rho _{3,l}^{\left( k \right)}}} + {\lambda _2}\frac{{\tan \phi  + j}}{2}{e^{j\angle {{\bf{s}}_\text{D}}\left[ l \right]}}{\bf{\tilde h}}_l^ * $.
By substituting ${{\bf{q}}_l}$ into (\ref{E4}), we obtain a closed form expression for $\lambda_2$, and hence
\begin{equation}
\begin{array}{l}
{{\bf{q}}_l} = {{\bf{s}}^{\left( k \right)}} - \frac{{{\boldsymbol{\mu }}_{{{\bf{q}}_l}}^{\left( k \right)}}}{{\rho _{3,l}^{\left( k \right)}}} + \\
\bigl\{ { - {\mathop{\rm Re}\nolimits} \bigl[ {\left( {\tan \phi  - j} \right){e^{ - j\angle {{\bf{s}}_\text{D}}\left[ l \right]}}{\bf{\tilde h}}_l^T\bigl( {{{\bf{s}}^{\left( k \right)}} - \frac{{{\boldsymbol{\mu }}_{{{\bf{q}}_l}}^{\left( k \right)}}}{{\rho _{3,l}^{\left( k \right)}}}} \bigr)} \bigr]+} \bigr.\\
{{\left. {  \sqrt {{\Gamma _u}{\sigma ^2}} \left| {{{\bf{s}}_\text{D}}\left[ l \right]} \right|\tan \phi } \right\}{e^{j\angle {{\bf{s}}_\text{D}}\left[ l \right]}}{\bf{\tilde h}}_l^ * } \mathord{\left/
 {\vphantom {{\left. { + \sqrt {{\Gamma _u}{\sigma ^2}} \left| {{{\bf{s}}_\text{D}}\left[ l \right]} \right|\tan \phi } \right\}{e^{j\angle {{\bf{s}}_\text{D}}\left[ l \right]}}{\bf{\tilde h}}_l^ * } {\left( {\tan \phi  - j} \right){\bf{\tilde h}}_l^H{{{\bf{\tilde h}}}_l}}}} \right.
 \kern-\nulldelimiterspace} {\left( {\tan \phi  - j} \right){\bf{\tilde h}}_l^H{{{\bf{\tilde h}}}_l}}}.
\end{array}
\end{equation}
Then we check whether the constraints (\ref{E7}) and (\ref{E8}) are satisfied. If so, it is the optimal solution. Otherwise, go to case 3.

\subsubsection{Case 3}
If ${\lambda _1} > 0$ and ${\lambda _2} = 0$, $ {{\bf{q}}_l}$ takes the form
${{\bf{q}}_l} = {{\bf{s}}^{\left( k \right)}} - \frac{{{\boldsymbol{\mu }}_{{{\bf{q}}_l}}^{\left( k \right)}}}{{\rho _{3,l}^{\left( k \right)}}} + {\lambda _1}\frac{{\tan \phi  - j}}{2}{e^{j\angle {{\bf{s}}_\text{D}}\left[ l \right]}}{\bf{\tilde h}}_l^ * $. By substituting ${{\bf{q}}_l}$ into (\ref{E3}), we obtain a closed form expression for $\lambda_1$, and hence
\begin{equation}
    \begin{array}{l}
{{\bf{q}}_l} = {{\bf{s}}^{\left( k \right)}} - \frac{{{\boldsymbol{\mu }}_{{{\bf{q}}_l}}^{\left( k \right)}}}{{\rho _{3,l}^{\left( k \right)}}} + \\
\bigl\{ { - {\mathop{\rm Re}\nolimits} \bigl[ {\left( {\tan \phi  + j} \right){e^{ - j\angle {{\bf{s}}_\text{D}}\left[ l \right]}}{\bf{\tilde h}}_l^T\bigl( {{{\bf{s}}^{\left( k \right)}} - \frac{{{\boldsymbol{\mu }}_{{{\bf{q}}_l}}^{\left( k \right)}}}{{\rho _{3,l}^{\left( k \right)}}}} \bigr)} \bigr] + } \bigr.\\
{{\left. {\sqrt {{\Gamma _u}{\sigma ^2}} \left| {{{\bf{s}}_\text{D}}\left[ l \right]} \right|\tan \phi } \right\}{e^{j\angle {{\bf{s}}_\text{D}}\left[ l \right]}}{\bf{\tilde h}}_l^ * } \mathord{\left/
 {\vphantom {{\left. {\sqrt {{\Gamma _u}{\sigma ^2}} \left| {{{\bf{s}}_\text{D}}\left[ l \right]} \right|\tan \phi } \right\}{e^{j\angle {{\bf{s}}_\text{D}}\left[ l \right]}}{\bf{\tilde h}}_l^ * } {\left( {\tan \phi  + j} \right){\bf{\tilde h}}_l^H{{{\bf{\tilde h}}}_l}}}} \right.
 \kern-\nulldelimiterspace} {\left( {\tan \phi  + j} \right){\bf{\tilde h}}_l^H{{{\bf{\tilde h}}}_l}}}.
\end{array}
\end{equation}
Then we check whether it satisfies the constraints (\ref{E7}) and (\ref{E8}). If so, it is optimal. Otherwise, go to case 4.

\subsubsection{Case 4}
If ${\lambda _1} > 0$ and ${\lambda _2} > 0$, substituting (\ref{E9}) into (\ref{E3}) and (\ref{E4}), we obtain 
\begin{equation}
\label{E16}
    {\lambda _1}\frac{{{{\tan }^2}\phi  + 1}}{2}{\bf{\tilde h}}_l^H{{\bf{\tilde h}}_l} + {\lambda _2}\frac{{{{\tan }^2}\phi  - 1}}{2}{\bf{\tilde h}}_l^H{{\bf{\tilde h}}_l} = {\zeta _1},
\end{equation}
\begin{equation}
\label{E18}
    {\lambda _1}\frac{{{{\tan }^2}\phi  - 1}}{2}{\bf{\tilde h}}_l^H{{\bf{\tilde h}}_l} + {\lambda _2}\frac{{{{\tan }^2}\phi  + 1}}{2}{\bf{\tilde h}}_l^H{{\bf{\tilde h}}_l} = {\zeta _2},
\end{equation}
where 
\begin{small}
\begin{equation}
    \begin{aligned}
{\zeta _1} =  &- {\mathop{\rm Re}\nolimits} \bigl[ {\left( {\tan \phi  + j} \right){e^{ - j\angle {{\bf{s}}_\text{D}}\left[ l \right]}}{\bf{\tilde h}}_l^T\bigl( {{{\bf{s}}^{\left( k \right)}} - \frac{{{\boldsymbol{\mu }}_{{{\bf{q}}_l}}^{\left( k \right)}}}{{\rho _{3,l}^{\left( k \right)}}}} \bigr)} \bigr]\\
 &+ \sqrt {{\Gamma _u}{\sigma ^2}} \left| {{{\bf{s}}_\text{D}}\left[ l \right]} \right|\tan \phi ,
\end{aligned}
\end{equation}
\end{small}
\begin{small}
\begin{equation}
    \begin{aligned}
{\zeta _2} =  &- {\mathop{\rm Re}\nolimits} \left[ {\left( {\tan \phi  - j} \right){e^{ - j\angle {{\bf{s}}_\text{D}}\left[ l \right]}}{\bf{\tilde h}}_l^T\left( {{{\bf{s}}^{\left( k \right)}} - \frac{{{\boldsymbol{\mu }}_{{{\bf{q}}_l}}^{\left( k \right)}}}{{\rho _{3,l}^{\left( k \right)}}}} \right)} \right]\\
 &+ \sqrt {{\Gamma _u}{\sigma ^2}} \left| {{{\bf{s}}_\text{D}}\left[ l \right]} \right|\tan \phi. 
\end{aligned}
\end{equation}
\end{small}
By solving the $2 \times 2$ linear system in (\ref{E16}) and (\ref{E18}), we can obtain closed form expressions for $\lambda_1$ and $\lambda_2$. The closed-form solution of ${{\bf{q}}_l}$ is obtained by substituting those expressions into (\ref{E9}), which yields
\begin{equation}
\begin{array}{l}
{{\bf{q}}_l} = {{\bf{s}}^{\left( k \right)}} - \frac{{{\boldsymbol{\mu }}_{{{\bf{q}}_l}}^{\left( k \right)}}}{{\rho _{3,l}^{\left( k \right)}}} + \\
\frac{{ - {e^{ - j\angle {{\bf{s}}_\text{D}}\left[ l \right]}}{\bf{\tilde h}}_l^T\left( {{{\bf{s}}^{\left( k \right)}} - \frac{{{\boldsymbol{\mu }}_{{{\bf{q}}_l}}^{\left( k \right)}}}{{\rho _{3,l}^{\left( k \right)}}}} \right) + \sqrt {{\Gamma _u}{\sigma ^2}} \left| {{{\bf{s}}_\text{D}}\left[ l \right]} \right|}}{{{\bf{\tilde h}}_l^H{{{\bf{\tilde h}}}_l}}}{e^{j\angle {{\bf{s}}_\text{D}}\left[ l \right]}}{\bf{\tilde h}}_l^ * .
\end{array}
\end{equation}

\subsection{Derivation of the Closed-form Solution of (45)}
\setcounter{equation}{0}
\renewcommand\theequation{D\arabic{equation}}

\subsubsection{Points A}
If ${{\bf{s}}_\text{D}}\left[ l \right]$ belongs to the set of points denoted by “A”, the optimization problem can be transformed to
\begin{equation}
    \begin{aligned}
\mathop {\min }\limits_{{{\bf{q}}_l}} \text{ }&{\bf{q}}_l^H{{\bf{q}}_l} - {\bf{q}}_l^H{{\bf{s}}^{\left( k \right)}} + \frac{{{\bf{q}}_l^H{\boldsymbol{\mu }}_{{{\bf{q}}_l}}^{\left( k \right)}}}{{\rho _{3,l}^{\left( k \right)}}} - {{\bf{s}}^{\left( k \right)H}}{{\bf{q}}_l} + \frac{{{\boldsymbol{\mu }}_{{{\bf{q}}_l}}^{\left( k \right)H}{{\bf{q}}_l}}}{{\rho _{3,l}^{\left( k \right)}}}\\
\text{s.t. } &{\bf{\tilde h}}_l^T{{\bf{q}}_l} = \sqrt {{\Gamma _u}{\sigma ^2}} {{\bf{s}}_\text{D}}\left[ l \right].
\end{aligned}
\end{equation}
The KKT conditions for thr problem include 
\begin{equation}
    {\bf{\tilde h}}_l^T{{\bf{q}}_l} - \sqrt {{\Gamma _u}{\sigma ^2}} {{\bf{s}}_\text{D}}\left[ l \right] = 0,
\end{equation}
\begin{equation}
    {{\bf{q}}_l} = {{\bf{s}}^{\left( k \right)}} - \frac{{{\boldsymbol{\mu }}_{{{\bf{q}}_l}}^{\left( k \right)}}}{{\rho _{3,l}^{\left( k \right)}}} - \nu {\bf{\tilde h}}_l^ *. 
\end{equation}
Hence, the closed-form solution is
\begin{small}
\begin{equation}
\begin{array}{l}
    {\bf{q}}_l^{\left( {k + 1} \right)} = {{\bf{s}}^{\left( k \right)}} - \frac{{{\boldsymbol{\mu }}_{{{\bf{q}}_l}}^{\left( k \right)}}}{{\rho _{3,l}^{\left( k \right)}}} - \frac{{{\bf{\tilde h}}_l^T{{\bf{s}}^{\left( k \right)}} - \frac{{{\bf{\tilde h}}_l^T{\boldsymbol{\mu }}_{{{\bf{q}}_l}}^{\left( k \right)}}}{{\rho _{3,l}^{\left( k \right)}}} - \sqrt {{\Gamma _u}{\sigma ^2}} {{\bf{s}}_\text{D}}\left[ l \right]}}{{{\bf{\tilde h}}_l^H{{{\bf{\tilde h}}}_l}}}{\bf{\tilde h}}_l^*.
    \end{array}
\end{equation}
\end{small}

\subsubsection{Points B}
When ${{\bf{s}}_\text{D}}\left[ l \right]$ belongs to the points denoted by “B”, we take the point in the first or second quadrant as an example. Points in the third or fourth quadrant can be handled in a similar way. The optimization problem can be written as
\begin{equation}
    \begin{aligned}
\mathop {\min }\limits_{{{\bf{q}}_l}} \text{ }&{\bf{q}}_l^H{{\bf{q}}_l} - {\bf{q}}_l^H{{\bf{s}}^{\left( k \right)}} + \frac{{{\bf{q}}_l^H{\boldsymbol{\mu }}_{{{\bf{q}}_l}}^{\left( k \right)}}}{{\rho _{3,l}^{\left( k \right)}}} - {{\bf{s}}^{\left( k \right)H}}{{\bf{q}}_l} + \frac{{{\boldsymbol{\mu }}_{{{\bf{q}}_l}}^{\left( k \right)H}{{\bf{q}}_l}}}{{\rho _{3,l}^{\left( k \right)}}}\\
\text{s.t. }&{\mathop{\rm Re}\nolimits} \left\{ {{\bf{\tilde h}}_l^T{{\bf{q}}_l}} \right\} - \sqrt {{\Gamma _u}{\sigma ^2}} {\mathop{\rm Re}\nolimits} \left\{ {{{\bf{s}}_\text{D}}\left[ l \right]} \right\} = 0,\\
& - {\mathop{\rm Im}\nolimits} \left\{ {{\bf{\tilde h}}_l^T{{\bf{q}}_l}} \right\} + \sqrt {{\Gamma _u}{\sigma ^2}} {\mathop{\rm Im}\nolimits} \left\{ {{{\bf{s}}_\text{D}}\left[ l \right]} \right\} \le 0.
\end{aligned}
\end{equation}
The KKT conditions include 
\begin{equation}
    {{\bf{q}}_l} = {{\bf{s}}^{\left( k \right)}} - \frac{{{\boldsymbol{\mu }}_{{{\bf{q}}_l}}^{\left( k \right)}}}{{\rho _{3,l}^{\left( k \right)}}} - \nu \frac{1}{2}{\bf{\tilde h}}_l^ *  + \lambda \frac{j}{2}{\bf{\tilde h}}_l^ * ,
\end{equation}
\begin{align}
    \frac{1}{2}{\bf{\tilde h}}_l^T{{\bf{q}}_l} + \frac{1}{2}{\bf{q}}_l^H{\bf{\tilde h}}_l^ *  - \sqrt {{\Gamma _u}{\sigma ^2}} {\mathop{\rm Re}\nolimits} \left\{ {{{\bf{s}}_\text{D}}\left[ l \right]} \right\} &= 0,
    \\
\label{F10}
    \frac{j}{2}{\bf{\tilde h}}_l^T{{\bf{q}}_l} - \frac{j}{2}{\bf{q}}_l^H{\bf{\tilde h}}_l^ *  + \sqrt {{\Gamma _u}{\sigma ^2}} {\mathop{\rm Im}\nolimits} \left\{ {{{\bf{s}}_\text{D}}\left[ l \right]} \right\} &\le 0,\\
    \lambda  &\ge 0,\\
    \lambda \Bigl( {\frac{j}{2}{\bf{\tilde h}}_l^T{{\bf{q}}_l} - \frac{j}{2}{\bf{q}}_l^H{\bf{\tilde h}}_l^ *  + \sqrt {{\Gamma _u}{\sigma ^2}} {\mathop{\rm Im}\nolimits} \left\{ {{{\bf{s}}_\text{D}}\left[ l \right]} \right\}} \Bigr) &= 0.
\end{align}
If $\lambda  = 0$, the solution can be written as
\vspace{-\baselineskip}
\begin{small}
\begin{equation}
\label{F14}
    {{\bf{q}}_l} = {{\bf{s}}^{\left( k \right)}} - \frac{{{\boldsymbol{\mu }}_{{{\bf{q}}_l}}^{\left( k \right)}}}{{\rho _{3,l}^{\left( k \right)}}} - \frac{{{\mathop{\rm Re}\nolimits} \left\{ {{\bf{\tilde h}}_l^T{{\bf{s}}^{\left( k \right)}} - \frac{{{\bf{\tilde h}}_l^T{\boldsymbol{\mu }}_{{{\bf{q}}_l}}^{\left( k \right)}}}{{\rho _{3,l}^{\left( k \right)}}} - \sqrt {{\Gamma _u}{\sigma ^2}} {{\bf{s}}_\text{D}}\left[ l \right]} \right\}}}{{{\bf{\tilde h}}_l^H{{{\bf{\tilde h}}}_l}}}{\bf{\tilde h}}_l^ *.
\end{equation}
\end{small}
\noindent If this solution satisfies (\ref{F10}), it is optimal. Otherwise, we assume $\lambda  > 0$, and we have
\begin{small}
\begin{equation}
\label{F16}
    {{\bf{q}}_l} = {{\bf{s}}^{\left( k \right)}} - \frac{{{\boldsymbol{\mu }}_{{{\bf{q}}_l}}^{\left( k \right)}}}{{\rho _{3,l}^{\left( k \right)}}} - \frac{{{\bf{\tilde h}}_l^T{{\bf{s}}^{\left( k \right)}} - \frac{{{\bf{\tilde h}}_l^T{\boldsymbol{\mu }}_{{{\bf{q}}_l}}^{\left( k \right)}}}{{\rho _{3,l}^{\left( k \right)}}} - \sqrt {{\Gamma _u}{\sigma ^2}} {{\bf{s}}_\text{D}}\left[ l \right]}}{{{\bf{\tilde h}}_l^H{{{\bf{\tilde h}}}_l}}}{\bf{\tilde h}}_l^ * .
\end{equation}
\end{small}

\subsubsection{Points C}
When ${{\bf{s}}_\text{D}}\left[ l \right]$ belongs to the points denoted by “C”, we take the point in the first quadrant as an example. The optimization problem can be written as

\vspace{-\baselineskip}
\begin{small}
\begin{equation}
    \begin{aligned}
\mathop {\min }\limits_{{{\bf{q}}_l}} \text{ }&{\bf{q}}_l^H{{\bf{q}}_l} - {\bf{q}}_l^H{{\bf{s}}^{\left( k \right)}} + \frac{{{\bf{q}}_l^H{\boldsymbol{\mu }}_{{{\bf{q}}_l}}^{\left( k \right)}}}{{\rho _{3,l}^{\left( k \right)}}} - {{\bf{s}}^{\left( k \right)H}}{{\bf{q}}_l} + \frac{{{\boldsymbol{\mu }}_{{{\bf{q}}_l}}^{\left( k \right)H}{{\bf{q}}_l}}}{{\rho _{3,l}^{\left( k \right)}}}\\
\text{s.t. }&- {\mathop{\rm Re}\nolimits} \left\{ {{\bf{\tilde h}}_l^T{{\bf{q}}_l}} \right\} + \sqrt {{\Gamma _u}{\sigma ^2}} {\mathop{\rm Re}\nolimits} \left\{ {{{\bf{s}}_\text{D}}\left[ l \right]} \right\} \le 0,\\
 &- {\mathop{\rm Im}\nolimits} \left\{ {{\bf{\tilde h}}_l^T{{\bf{q}}_l}} \right\} + \sqrt {{\Gamma _u}{\sigma ^2}} {\mathop{\rm Im}\nolimits} \left\{ {{{\bf{s}}_\text{D}}\left[ l \right]} \right\} \le 0.
\end{aligned}
\end{equation}
\end{small}
The KKT conditions include 
\begin{equation}
    {{\bf{q}}_l} = {{\bf{s}}^{\left( k \right)}} - \frac{{{\boldsymbol{\mu }}_{{{\bf{q}}_l}}^{\left( k \right)}}}{{\rho _{3,l}^{\left( k \right)}}} + {\lambda _1}\frac{1}{2}{\bf{\tilde h}}_l^ *  + {\lambda _2}\frac{j}{2}{\bf{\tilde h}}_l^ *, 
\end{equation}
\begin{align}
\label{F21}
     - \frac{1}{2}{\bf{\tilde h}}_l^T{{\bf{q}}_l} - \frac{1}{2}{\bf{q}}_l^H{\bf{\tilde h}}_l^ *  + \sqrt {{\Gamma _u}{\sigma ^2}} {\mathop{\rm Re}\nolimits} \left\{ {{{\bf{s}}_\text{D}}\left[ l \right]} \right\} &\le 0,\\
\label{F22}
    \frac{j}{2}{\bf{\tilde h}}_l^T{{\bf{q}}_l} - \frac{j}{2}{\bf{q}}_l^H{\bf{\tilde h}}_l^ *  + \sqrt {{\Gamma _u}{\sigma ^2}} {\mathop{\rm Im}\nolimits} \left\{ {{{\bf{s}}_\text{D}}\left[ l \right]} \right\} &\le 0,\\
    {\lambda _1} \ge 0, {\lambda _2} &\ge 0,\\
    {\lambda _1}\left( { - \frac{1}{2}{\bf{\tilde h}}_l^T{{\bf{q}}_l} - \frac{1}{2}{\bf{q}}_l^H{\bf{\tilde h}}_l^ *  + \sqrt {{\Gamma _u}{\sigma ^2}} {\mathop{\rm Re}\nolimits} \left\{ {{{\bf{s}}_\text{D}}\left[ l \right]} \right\}} \right) &= 0,\\
    {\lambda _2}\left( {\frac{j}{2}{\bf{\tilde h}}_l^T{{\bf{q}}_l} - \frac{j}{2}{\bf{q}}_l^H{\bf{\tilde h}}_l^ *  + \sqrt {{\Gamma _u}{\sigma ^2}} {\mathop{\rm Im}\nolimits} \left\{ {{{\bf{s}}_\text{D}}\left[ l \right]} \right\}} \right) &= 0.
\end{align}
If ${\lambda _1} = 0$ and ${\lambda _2} = 0$, we check whether ${{\bf{q}}_l} = {{\bf{s}}^{\left( k \right)}} - \frac{{{\boldsymbol{\mu }}_{{{\bf{q}}_l}}^{\left( k \right)}}}{{\rho _{3,l}^{\left( k \right)}}}$ satisfies (\ref{F21}) and (\ref{F22}). If so, it is optimal. Otherwise, we move on to the assumption that ${\lambda _1} = 0$ and ${\lambda _2} > 0$. With this assumption, ${{\bf{q}}_l}$ can be written as
\begin{small}
\begin{equation}
\label{F28}
    {{\bf{q}}_l} = {{\bf{s}}^{\left( k \right)}} - \frac{{{\boldsymbol{\mu }}_{{{\bf{q}}_l}}^{\left( k \right)}}}{{\rho _{3,l}^{\left( k \right)}}} - j\frac{{{\mathop{\rm Im}\nolimits} \left\{ {{\bf{\tilde h}}_l^T{{\bf{s}}^{\left( k \right)}} - \frac{{{\bf{\tilde h}}_l^T{\boldsymbol{\mu }}_{{{\bf{q}}_l}}^{\left( k \right)}}}{{\rho _{3,l}^{\left( k \right)}}} - \sqrt {{\Gamma _u}{\sigma ^2}} {{\bf{s}}_\text{D}}\left[ l \right]} \right\}}}{{{\bf{\tilde h}}_l^H{{{\bf{\tilde h}}}_l}}}{\bf{\tilde h}}_l^ *. 
\end{equation}
\end{small}
\noindent We check whether it satisfies (\ref{F21}). If so, it is optimal. Otherwise, we move on to the assumption that ${\lambda _1} > 0$ and ${\lambda _2} = 0$. With the assumption, ${{\bf{q}}_l}$ can be written as
\begin{equation}
\label{F30}
    {{\bf{q}}_l} = {{\bf{s}}^{\left( k \right)}} - \frac{{{\boldsymbol{\mu }}_{{{\bf{q}}_l}}^{\left( k \right)}}}{{\rho _{3,l}^{\left( k \right)}}} - \frac{{{\mathop{\rm Re}\nolimits} \left\{ {{\bf{\tilde h}}_l^T{{\bf{s}}^{\left( k \right)}} - \frac{{{\bf{\tilde h}}_l^T{\boldsymbol{\mu }}_{{{\bf{q}}_l}}^{\left( k \right)}}}{{\rho _{3,l}^{\left( k \right)}}} - \sqrt {{\Gamma _u}{\sigma ^2}} {{\bf{s}}_\text{D}}\left[ l \right]} \right\}}}{{{\bf{\tilde h}}_l^H{{{\bf{\tilde h}}}_l}}}{\bf{\tilde h}}_l^ *. 
\end{equation}
We check whether it satisfies (\ref{F22}). If so, it is optimal. Otherwise, we move on to the assumption that ${\lambda _1} > 0$ and ${\lambda _2} > 0$. Based on this assumption, we have
\begin{equation}
    \left\{ {\begin{array}{*{20}{c}}
{{{\bf{q}}_l} = {{\bf{s}}^{\left( k \right)}} - \frac{{{\boldsymbol{\mu }}_{{{\bf{q}}_l}}^{\left( k \right)}}}{{\rho _{3,l}^{\left( k \right)}}} + {\lambda _1}\frac{1}{2}{\bf{\tilde h}}_l^ *  + {\lambda _2}\frac{j}{2}{\bf{\tilde h}}_l^ * },\\
{ - \frac{1}{2}{\bf{\tilde h}}_l^T{{\bf{q}}_l} - \frac{1}{2}{\bf{q}}_l^H{\bf{\tilde h}}_l^ *  + \sqrt {{\Gamma _u}{\sigma ^2}} {\mathop{\rm Re}\nolimits} \left\{ {{{\bf{s}}_\text{D}}\left[ l \right]} \right\} = 0},\\
{\frac{j}{2}{\bf{\tilde h}}_l^T{{\bf{q}}_l} - \frac{j}{2}{\bf{q}}_l^H{\bf{\tilde h}}_l^ *  + \sqrt {{\Gamma _u}{\sigma ^2}} {\mathop{\rm Im}\nolimits} \left\{ {{{\bf{s}}_\text{D}}\left[ l \right]} \right\} = 0}.
\end{array}} \right.
\end{equation}
According to the equations, ${{\bf{q}}_l}$ can be written as
\begin{equation}
\label{F32}
    {{\bf{q}}_l} = {{\bf{s}}^{\left( k \right)}} - \frac{{{\boldsymbol{\mu }}_{{{\bf{q}}_l}}^{\left( k \right)}}}{{\rho _{3,l}^{\left( k \right)}}} - \frac{{{\bf{\tilde h}}_l^T{{\bf{s}}^{\left( k \right)}} - \frac{{{\bf{\tilde h}}_l^T{\boldsymbol{\mu }}_{{{\bf{q}}_l}}^{\left( k \right)}}}{{\rho _{3,l}^{\left( k \right)}}} - \sqrt {{\Gamma _u}{\sigma ^2}} {{\bf{s}}_\text{D}}\left[ l \right]}}{{{\bf{\tilde h}}_l^H{{{\bf{\tilde h}}}_l}}}{\bf{\tilde h}}_l^ *. 
\end{equation}

When ${{\bf{s}}_\text{D}}\left[ l \right]$ is in the other three quadrants, the problem can be handled in a similar way. Note that the order in which we check the four cases does not affect the result.

\subsubsection{Points D}
When ${{\bf{s}}_\text{D}}\left[ l \right]$ belongs to the points denoted by “D”, we take the point in the first or fourth quadrant as an example. The optimization problem can be written as 
\begin{small}
\begin{equation}
    \begin{aligned}
\mathop {\min }\limits_{{{\bf{q}}_l}} \text{ }&{\bf{q}}_l^H{{\bf{q}}_l} - {\bf{q}}_l^H{{\bf{s}}^{\left( k \right)}} + \frac{{{\bf{q}}_l^H{\boldsymbol{\mu }}_{{{\bf{q}}_l}}^{\left( k \right)}}}{{\rho _{3,l}^{\left( k \right)}}} - {{\bf{s}}^{\left( k \right)H}}{{\bf{q}}_l} + \frac{{{\boldsymbol{\mu }}_{{{\bf{q}}_l}}^{\left( k \right)H}{{\bf{q}}_l}}}{{\rho _{3,l}^{\left( k \right)}}}\\
\text{s.t. }&- {\mathop{\rm Re}\nolimits} \left\{ {{\bf{\tilde h}}_l^T{{\bf{q}}_l}} \right\} + \sqrt {{\Gamma _u}{\sigma ^2}} {\mathop{\rm Re}\nolimits} \left\{ {{{\bf{s}}_\text{D}}\left[ l \right]} \right\} \le 0,\\
&{\mathop{\rm Im}\nolimits} \left\{ {{\bf{\tilde h}}_l^T{{\bf{q}}_l}} \right\} - \sqrt {{\Gamma _u}{\sigma ^2}} {\mathop{\rm Im}\nolimits} \left\{ {{{\bf{s}}_\text{D}}\left[ l \right]} \right\} = 0.
\end{aligned}
\end{equation}
\end{small}
\noindent The KKT conditions include
\begin{equation}
    {{\bf{q}}_l} = {{\bf{s}}^{\left( k \right)}} - \frac{{{\boldsymbol{\mu }}_{{{\bf{q}}_l}}^{\left( k \right)}}}{{\rho _{3,l}^{\left( k \right)}}} + \lambda \frac{1}{2}{\bf{\tilde h}}_l^ *  - \nu \frac{j}{2}{\bf{\tilde h}}_l^ *, 
\end{equation}
\begin{equation}
\label{F39}
     - \frac{1}{2}{\bf{\tilde h}}_l^T{{\bf{q}}_l} - \frac{1}{2}{\bf{q}}_l^H{\bf{\tilde h}}_l^ *  + \sqrt {{\Gamma _u}{\sigma ^2}} {\mathop{\rm Re}\nolimits} \left\{ {{{\bf{s}}_\text{D}}\left[ l \right]} \right\} \le 0,
\end{equation}
\begin{equation}
     - \frac{j}{2}{\bf{\tilde h}}_l^T{{\bf{q}}_l} + \frac{j}{2}{\bf{q}}_l^H{\bf{\tilde h}}_l^ *  - \sqrt {{\Gamma _u}{\sigma ^2}} {\mathop{\rm Im}\nolimits} \left\{ {{{\bf{s}}_\text{D}}\left[ l \right]} \right\} = 0,
\end{equation}
\begin{equation}
    \lambda  \ge 0,
\end{equation}
\begin{equation}
    \lambda \left( { - \frac{1}{2}{\bf{\tilde h}}_l^T{{\bf{q}}_l} - \frac{1}{2}{\bf{q}}_l^H{\bf{\tilde h}}_l^ *  + \sqrt {{\Gamma _u}{\sigma ^2}} {\mathop{\rm Re}\nolimits} \left\{ {{{\bf{s}}_\text{D}}\left[ l \right]} \right\}} \right) = 0.
\end{equation}

If $\lambda  = 0$, the solution can be written as
\begin{small}
\begin{equation}
\begin{aligned}
\label{F44}
    {{\bf{q}}_l} = &{{\bf{s}}^{\left( k \right)}} - \frac{{{\boldsymbol{\mu }}_{{{\bf{q}}_l}}^{\left( k \right)}}}{{\rho _{3,l}^{\left( k \right)}}} \\
    &- j\frac{{{\mathop{\rm Im}\nolimits} \left\{ {{\bf{\tilde h}}_l^T{{\bf{s}}^{\left( k \right)}} - \frac{{{\bf{\tilde h}}_l^T{\boldsymbol{\mu }}_{{{\bf{q}}_l}}^{\left( k \right)}}}{{\rho _{3,l}^{\left( k \right)}}} - \sqrt {{\Gamma _u}{\sigma ^2}} {{\bf{s}}_\text{D}}\left[ l \right]} \right\}}}{{{\bf{\tilde h}}_l^H{{{\bf{\tilde h}}}_l}}}{\bf{\tilde h}}_l^ * .
\end{aligned}
\end{equation}
\end{small}
If (\ref{F44}) satisfies (\ref{F39}), it is optimal. Otherwise, we assume $\lambda  > 0$, then we have
\begin{small}
\begin{equation}
\label{F46}
    {{\bf{q}}_l} = {{\bf{s}}^{\left( k \right)}} - \frac{{{\boldsymbol{\mu }}_{{{\bf{q}}_l}}^{\left( k \right)}}}{{\rho _{3,l}^{\left( k \right)}}} - \frac{{{\bf{\tilde h}}_l^T{{\bf{s}}^{\left( k \right)}} - \frac{{{\bf{\tilde h}}_l^T{\boldsymbol{\mu }}_{{{\bf{q}}_l}}^{\left( k \right)}}}{{\rho _{3,l}^{\left( k \right)}}} - \sqrt {{\Gamma _u}{\sigma ^2}} {{\bf{s}}_\text{D}}\left[ l \right]}}{{{\bf{\tilde h}}_l^H{{{\bf{\tilde h}}}_l}}}{\bf{\tilde h}}_l^ * .
\end{equation}
\end{small}
In the case of points “D” in the second and third quadrant, the optimization problem can be handled in a similar way.

\subsection{Convergence Analysis of SCA}
\setcounter{equation}{0}
\renewcommand\theequation{E\arabic{equation}}

\subsubsection{Proof of Property 1}
Assume that $\left\{ {{{\bf{s}}^{\left( j \right)}},{t^{\left( j \right)}}} \right\}$ is a feasible point of (28). Hence, we have
\begin{equation}
    {{\bf{s}}^{\left( j \right)H}}{{\boldsymbol{\Psi }}_\text{SL}}{{\bf{s}}^{\left( j \right)}} - \frac{{{{\bf{s}}^{\left( j \right)H}}{{\boldsymbol{\Psi }}_\text{ML}}{{\bf{s}}^{\left( j \right)}}}}{{{t^{\left( j \right)}}}} \le 0,
\end{equation}
\begin{equation}
     - {{\bf{s}}^{\left( j \right)H}}{{\bf{e}}_m}{\bf{e}}_m^T{{\bf{s}}^{\left( j \right)}} + \left( {1 - \varepsilon } \right){P_0}/{N_\text{t}} \le 0,m = 1, \cdots ,{N_\text{s}}{N_\text{t}}.
\end{equation}
According to the derivation of (32), we have
\begin{small}
\begin{equation}
    {{\bf{s}}^{\left( j \right)H}}{{\boldsymbol{\Psi }}_\text{SL}}{{\bf{s}}^{\left( j \right)}} - \frac{{2{\mathop{\rm Re}\nolimits} \left\{ {{{\bf{s}}^{\left( j \right)H}}{{\boldsymbol{\Psi }}_\text{ML}}{{\bf{s}}^{\left( j \right)}}} \right\}}}{{{{\bar t}^{\left( j \right)}}}} + \frac{{{{\bf{s}}^{\left( j \right)H}}{{\boldsymbol{\Psi }}_\text{ML}}{{\bf{s}}^{\left( j \right)}}}}{{{{\bar t}^{\left( j \right)2}}}}{t^{\left( j \right)}} \le 0.
\end{equation}
\end{small}
Since the nonconvex part in constraint (22) is replaced by its first-order Taylor expansion, then we can obtain that
\begin{small}
\begin{equation}
    \begin{array}{l}
{{\bf{s}}^{\left( j \right)H}}{{\bf{e}}_m}{\bf{e}}_m^T{{\bf{s}}^{\left( j \right)}} - 2{\mathop{\rm Re}\nolimits} \left\{ {{{\bf{s}}^{\left( j \right)H}}{{\bf{e}}_m}{\bf{e}}_m^T{{\bf{s}}^{\left( j \right)}}} \right\} + \left( {1 - \varepsilon } \right){P_0}/{N_\text{t}} \le 0,\\
m = 1, \cdots ,{N_\text{s}}{N_\text{t}}.
\end{array}
\end{equation}
\end{small}

Hence, $\left\{ {{{\bf{s}}^{\left( j \right)}},{t^{\left( j \right)}}} \right\}$ is also a feasible point of (33). Suppose that $\left\{ {{{{\bf{\hat s}}}^{\left( {j + 1} \right)}},{{\hat t}^{\left( {j + 1} \right)}}} \right\}$ is the optimal solution of the $\left( {j + 1} \right)$-th iteration of SCA. Since the solution to (30) is updated by $\left\{ {{{\bf{s}}^{\left( {j + 1} \right)}},{t^{\left( {j + 1} \right)}}} \right\} = \left\{ {{{{\bf{\hat s}}}^{\left( {j + 1} \right)}},{{\hat t}^{\left( {j + 1} \right)}}} \right\}$, we have
\begin{equation}
    \begin{aligned}
&{{\bf{s}}^{\left( {j + 1} \right)H}}{{\boldsymbol{\Psi }}_\text{SL}}{{\bf{s}}^{\left( {j + 1} \right)}} - \frac{{2{\mathop{\rm Re}\nolimits} \left\{ {{{\bf{s}}^{\left( j \right)H}}{{\boldsymbol{\Psi }}_\text{ML}}{{\bf{s}}^{\left( {j + 1} \right)}}} \right\}}}{{{{\bar t}^{\left( j \right)}}}}\\
 &+ \frac{{{{\bf{s}}^{\left( j \right)H}}{{\boldsymbol{\Psi }}_\text{ML}}{{\bf{s}}^{\left( j \right)}}}}{{{{\bar t}^{\left( j \right)2}}}}{t^{\left( {j + 1} \right)}} \le 0,
\end{aligned}
\end{equation}
and
\begin{equation}
    \begin{array}{l}
{{\bf{s}}^{\left( j \right)H}}{{\bf{e}}_m}{\bf{e}}_m^T{{\bf{s}}^{\left( j \right)}} - 2{\mathop{\rm Re}\nolimits} \left\{ {{{\bf{s}}^{\left( j \right)H}}{{\bf{e}}_m}{\bf{e}}_m^T{{\bf{s}}^{\left( {j + 1} \right)}}} \right\}\\
 + \left( {1 - \varepsilon } \right){P_0}/{N_\text{t}} \le 0,m = 1, \cdots ,{N_\text{s}}{N_\text{t}}.
\end{array}
\end{equation}

As the nonconvex constraints are approximated by their convex upper bound functions, we have that: 
\begin{equation}
\begin{aligned}
&{{\bf{s}}^{\left( {j + 1} \right)H}}{{\boldsymbol{\Psi }}_\text{SL}}{{\bf{s}}^{\left( {j + 1} \right)}} - \frac{{{{\bf{s}}^{\left( {j + 1} \right)H}}{{\boldsymbol{\Psi }}_\text{ML}}{{\bf{s}}^{\left( {j + 1} \right)}}}}{{{t^{\left( {j + 1} \right)}}}}\\
 \le &{{\bf{s}}^{\left( {j + 1} \right)H}}{{\boldsymbol{\Psi }}_\text{SL}}{{\bf{s}}^{\left( {j + 1} \right)}} - \frac{{2{\mathop{\rm Re}\nolimits} \left\{ {{{\bf{s}}^{\left( j \right)H}}{{\boldsymbol{\Psi }}_\text{ML}}{{\bf{s}}^{\left( {j + 1} \right)}}} \right\}}}{{{{\bar t}^{\left( j \right)}}}}\\
 &+ \frac{{{{\bf{s}}^{\left( j \right)H}}{{\boldsymbol{\Psi }}_\text{ML}}{{\bf{s}}^{\left( j \right)}}}}{{{{\bar t}^{\left( j \right)2}}}}{t^{\left( {j + 1} \right)}}\\
 \le &0,
\end{aligned}
\end{equation}
and
\begin{equation}
    \begin{array}{l}
 - {{\bf{s}}^{\left( {j + 1} \right)H}}{{\bf{e}}_m}{\bf{e}}_m^T{{\bf{s}}^{\left( {j + 1} \right)}} + \left( {1 - \varepsilon } \right){P_0}/{N_\text{t}}\\
 \le {{\bf{s}}^{\left( j \right)H}}{{\bf{e}}_m}{\bf{e}}_m^T{{\bf{s}}^{\left( j \right)}} - 2{\mathop{\rm Re}\nolimits} \left\{ {{{\bf{s}}^{\left( j \right)H}}{{\bf{e}}_m}{\bf{e}}_m^T{{\bf{s}}^{\left( {j + 1} \right)}}} \right\} + \left( {1 - \varepsilon } \right){\frac{{{P_0}}}{{{N_{\rm{t}}}}}}\\
 \le 0.
\end{array}
\end{equation}
Therefore, $\left\{ {{{\bf{s}}^{\left( {j + 1} \right)}},{t^{\left( {j + 1} \right)}}} \right\}$ is a feasible point of (28).

\subsubsection{Proof of Property 2}
Based on the definition of the objective function, we have
\begin{equation}
    \begin{aligned}
&f\left( {{{\bf{s}}^{\left( {j + 1} \right)}},{t^{\left( {j + 1} \right)}}} \right) =- {t^{\left( {j + 1} \right)}} + {\tilde\lambda _{\bf{G}}}{{\bf{s}}^{\left( {j + 1} \right)H}}{\bf{G}}_{{\rm{side}}}^{\left( {i + 1} \right)}{{\bf{s}}^{\left( {j + 1} \right)}}\\
 \le  &- {t^{\left( {j + 1} \right)}} + {\tilde\lambda _{\bf{G}}}{{\bf{s}}^{\left( {j + 1} \right)H}}{\bf{G}}_{{\rm{side}}}^{\left( {i + 1} \right)}{{\bf{s}}^{\left( {j + 1} \right)}} + {\lambda _{\bf{s}}}\left\| {{{\bf{s}}^{\left( {j + 1} \right)}} - {{\bf{s}}^{\left( j \right)}}} \right\|_2^2\\
 \le & - {t^{\left( j \right)}} + {\tilde\lambda _{\bf{G}}}{{\bf{s}}^{\left( j \right)H}}{\bf{G}}_{{\rm{side}}}^{\left( {i + 1} \right)}{{\bf{s}}^{\left( j \right)}} + {\lambda _{\bf{s}}}\left\| {{{\bf{s}}^{\left( j \right)}} - {{\bf{s}}^{\left( j \right)}}} \right\|_2^2\\
 =  &- {t^{\left( j \right)}} + {\tilde\lambda _{\bf{G}}}{{\bf{s}}^{\left( j \right)H}}{\bf{G}}_{{\rm{side}}}^{\left( {i + 1} \right)}{{\bf{s}}^{\left( j \right)}}\\
 \le &f\left( {{{\bf{s}}^{\left( j \right)}},{t^{\left( j \right)}}} \right).
\end{aligned}
\end{equation}

The second inequality holds because $\left\{ {{{\bf{s}}^{\left( {j + 1} \right)}},{t^{\left( {j + 1} \right)}}} \right\}$ is the optimal solution of (33). Therefore, we can reach the conclusion that $\left\{ {f\left( {{{\bf{s}}^{\left( j \right)}},{t^{\left( j \right)}}} \right)} \right\}_{j = 0}^\infty $ is non-increasing.

Since ${\bf{G}}_{{\rm{side}}}^{\left( {i + 1} \right)}$, ${{\boldsymbol{\Psi }}_\text{ML}}$ and ${{\boldsymbol{\Psi }}_\text{SL}}$ are positive definite, we have
\begin{equation}
    \begin{aligned}
f\left( {{\bf{s}},t} \right) =& - t + {\tilde\lambda _{\bf{G}}}{{\bf{s}}^H}{\bf{G}}_{{\rm{side}}}^{\left( {i + 1} \right)}{\bf{s}}\\
 \ge & - \frac{{{{\bf{s}}^H}{{\boldsymbol{\Psi }}_\text{ML}}{\bf{s}}}}{{{{\bf{s}}^H}{{\boldsymbol{\Psi }}_\text{SL}}{\bf{s}}}} + {\tilde\lambda _{\bf{G}}}{{\bf{s}}^H}{\bf{G}}_{{\rm{side}}}^{\left( {i + 1} \right)}{\bf{s}}\\
 \ge & - \frac{{{{\bf{s}}^H}{{\boldsymbol{\Psi }}_\text{ML}}{\bf{s}}}}{{{{\bf{s}}^H}{{\boldsymbol{\Psi }}_\text{SL}}{\bf{s}}}}\\
 \ge & - {\lambda _{\max }},
\end{aligned}
\end{equation}
where ${\lambda _{\max }}$ represents the maximum generalized eigenvalue of ${{\boldsymbol{\Psi }}_\text{ML}}$ and ${{\boldsymbol{\Psi }}_\text{SL}}$. Hence, $\left\{ {f\left( {{{\bf{s}}^{\left( j \right)}},{t^{\left( j \right)}}} \right)} \right\}_{j = 0}^\infty $ has a lower bound. Since the sequence is monotonically non-increasing and lower-bounded, it converges to a finite value.

\subsubsection{Proof of Property 3}
According to the first two properties, the proof of Property 3 can be established from the proof of Theorem 1 in Appendix A of [r2].

\subsection{Convergence Analysis of AO}
\setcounter{equation}{0}
\renewcommand\theequation{F\arabic{equation}}
First, we prove that $\left\{ {g\left( {{{\bf{g}}^{\left( i \right)}},{{\bf{s}}^{\left( i \right)}},{t^{\left( i \right)}}} \right)} \right\}_{i = 0}^\infty $ is non-increasing. According to (28), we have

\vspace{-\baselineskip}
\begin{small}
\begin{equation}
    \begin{aligned}
&g\left( {{{\bf{g}}^{\left( {i + 1} \right)}},{{\bf{s}}^{\left( {i + 1} \right)}},{t^{\left( {i + 1} \right)}}} \right)\\
 =  &- {t^{\left( {i + 1} \right)}} + {\tilde\lambda _{\bf{G}}}\sum\limits_{n = 1,n \ne {N_\text{s}} + {N_\text{cp}}}^{2\left( {{N_\text{s}} + {N_\text{cp}}} \right) - 1} {{{\left| {{{\bf{g}}^{\left( {i + 1} \right)H}}{{{\bf{\bar G}}}_n}{{\bf{s}}^{\left( {i + 1} \right)}}} \right|}^2}}  + \\
&{\tilde\lambda _{\bf{G}}}\sum\limits_{n = {N_\text{s}} + {N_\text{cp}}}^{2\left( {{N_\text{s}} + {N_\text{cp}}} \right) - 1} {{{\left| {{{\bf{g}}^{\left( {i + 1} \right)H}}{{{\bf{\bar G}}}_n}{{\bf{s}}_{{\rm{pre}}}}} \right|}^2}}  + {\tilde\lambda _{\bf{G}}}\sum\limits_{n = 1}^{{N_\text{s}} + {N_\text{cp}}} {{{\left| {{{\bf{g}}^{\left( {i + 1} \right)H}}{{{\bf{\bar G}}}_n}{{\bf{s}}_{{\rm{post}}}}} \right|}^2}} \\
 = & - {t^{\left( {i + 1} \right)}} + {\tilde\lambda _{\bf{G}}}{{\bf{s}}^{\left( {i + 1} \right)H}}{\bf{G}}_{{\rm{side}}}^{\left( {i + 1} \right)}{{\bf{s}}^{\left( {i + 1} \right)}} + \\
&{\tilde\lambda _{\bf{G}}}\sum\limits_{n = {N_\text{s}} + {N_\text{cp}}}^{2\left( {{N_\text{s}} + {N_\text{cp}}} \right) - 1} {{{\left| {{{\bf{g}}^{\left( {i + 1} \right)H}}{{{\bf{\bar G}}}_n}{{\bf{s}}_{{\rm{pre}}}}} \right|}^2}}  + {\tilde\lambda _{\bf{G}}}\sum\limits_{n = 1}^{{N_\text{s}} + {N_\text{cp}}} {{{\left| {{{\bf{g}}^{\left( {i + 1} \right)H}}{{{\bf{\bar G}}}_n}{{\bf{s}}_{{\rm{post}}}}} \right|}^2}} \\
 \le  &- {t^{\left( i \right)}} + {\tilde\lambda _{\bf{G}}}{{\bf{s}}^{\left( i \right)H}}{\bf{G}}_{{\rm{side}}}^{\left( {i + 1} \right)}{{\bf{s}}^{\left( i \right)}} + \\
&{\tilde\lambda _{\bf{G}}}\sum\limits_{n = {N_\text{s}} + {N_\text{cp}}}^{2\left( {{N_\text{s}} + {N_\text{cp}}} \right) - 1} {{{\left| {{{\bf{g}}^{\left( {i + 1} \right)H}}{{{\bf{\bar G}}}_n}{{\bf{s}}_{{\rm{pre}}}}} \right|}^2}}  + {\tilde\lambda _{\bf{G}}}\sum\limits_{n = 1}^{{N_\text{s}} + {N_\text{cp}}} {{{\left| {{{\bf{g}}^{\left( {i + 1} \right)H}}{{{\bf{\bar G}}}_n}{{\bf{s}}_{{\rm{post}}}}} \right|}^2}} \\
 =  &- {t^{\left( i \right)}} + {\tilde\lambda _{\bf{G}}}{{\bf{g}}^{\left( {i + 1} \right)H}}{{\bf{D}}^{\left( i \right)}}{{\bf{g}}^{\left( {i + 1} \right)}}\\
 \le & - {t^{\left( i \right)}} + {\tilde\lambda _{\bf{G}}}{{\bf{g}}^{\left( i \right)H}}{{\bf{D}}^{\left( i \right)}}{{\bf{g}}^{\left( i \right)}}\\
 = &g\left( {{{\bf{g}}^{\left( i \right)}},{{\bf{s}}^{\left( i \right)}},{t^{\left( i \right)}}} \right).
\end{aligned}
\end{equation}
\end{small}

The first constraint holds because $\left\{ {{{\bf{s}}^{\left( {i + 1} \right)}},{t^{\left( {i + 1} \right)}}} \right\}$ is the optimal solution of (30). The second constraint holds because ${{\bf{g}}^{\left( {i + 1} \right)}}$ is the optimal solution of (29).

Next, we prove that $\left\{ {g\left( {{{\bf{g}}^{\left( i \right)}},{{\bf{s}}^{\left( i \right)}},{t^{\left( i \right)}}} \right)} \right\}_{i = 0}^\infty $ has a lower bound: 
\begin{equation}
    \begin{aligned}
&g\left( {{\bf{g}},{\bf{s}},t} \right)\\
 = & - t + {\tilde\lambda _{\bf{G}}}\sum\limits_{n = 1,n \ne {N_\text{s}} + {N_\text{cp}}}^{2\left( {{N_\text{s}} + {N_\text{cp}}} \right) - 1} {{{\left| {{{\bf{g}}^H}{{{\bf{\bar G}}}_n}{\bf{s}}} \right|}^2}}  + \\
&{\tilde\lambda _{\bf{G}}}\sum\limits_{n = {N_\text{s}} + {N_\text{cp}}}^{2\left( {{N_\text{s}} + {N_\text{cp}}} \right) - 1} {{{\left| {{{\bf{g}}^H}{{{\bf{\bar G}}}_n}{{\bf{s}}_{{\rm{pre}}}}} \right|}^2}}  + {\tilde\lambda _{\bf{G}}}\sum\limits_{n = 1}^{{N_\text{s}} + {N_\text{cp}}} {{{\left| {{{\bf{g}}^H}{{{\bf{\bar G}}}_n}{{\bf{s}}_{{\rm{post}}}}} \right|}^2}} \\
 \ge  &- \frac{{{{\bf{s}}^H}{{\boldsymbol{\Psi }}_\text{ML}}{\bf{s}}}}{{{{\bf{s}}^H}{{\boldsymbol{\Psi }}_\text{SL}}{\bf{s}}}} + {\tilde\lambda _{\bf{G}}}\sum\limits_{n = 1,n \ne {N_\text{s}} + {N_\text{cp}}}^{2\left( {{N_\text{s}} + {N_\text{cp}}} \right) - 1} {{{\left| {{{\bf{g}}^H}{{{\bf{\bar G}}}_n}{\bf{s}}} \right|}^2}}  + \\
&{\tilde\lambda _{\bf{G}}}\sum\limits_{n = {N_\text{s}} + {N_\text{cp}}}^{2\left( {{N_\text{s}} + {N_\text{cp}}} \right) - 1} {{{\left| {{{\bf{g}}^H}{{{\bf{\bar G}}}_n}{{\bf{s}}_{{\rm{pre}}}}} \right|}^2}}  + {\tilde\lambda _{\bf{G}}}\sum\limits_{n = 1}^{{N_\text{s}} + {N_\text{cp}}} {{{\left| {{{\bf{g}}^H}{{{\bf{\bar G}}}_n}{{\bf{s}}_{{\rm{post}}}}} \right|}^2}} \\
 \ge  &- \frac{{{{\bf{s}}^H}{{\boldsymbol{\Psi }}_\text{ML}}{\bf{s}}}}{{{{\bf{s}}^H}{{\boldsymbol{\Psi }}_\text{SL}}{\bf{s}}}}\\
 \ge & - {\lambda _{\max }}.
\end{aligned}
\end{equation}

Since the sequence is non-increasing and lower-bounded, it converges to a finite value.

\end{document}